\documentclass[twocolumn,psfig,aps,showpacs]{revtex4}
\usepackage{amsfonts}
\usepackage{amsmath}
\usepackage{graphicx}
\usepackage{bm}
\usepackage{amssymb}
\usepackage{dcolumn}

\begin{document}

\title{Possible d$^0$ ferromagnetism in MgO doped with nitrogen}

\author{Bo Gu$^{1}$, Nejat Bulut$^{1,2}$, Timothy Ziman$^{3}$, and Sadamichi
Maekawa$^{1,2}$}

\affiliation{$^1$ Institute for Materials Research, Tohoku
University, Sendai 980-8577, Japan\\ $^{2}$JST, CREST, 3-Sanbancho,
Chiyoda-ku, Tokyo 102-0075, Japan\\$^{3}$C.N.R.S. and Institut Laue Langevin,
Bo\^\i te Postale 156, F-38042 Grenoble Cedex 9, France}

\begin{abstract}
We study the possibility of  d$^0$ ferromagnetism in the compound
MgO doped with nitrogen (N). The Haldane-Anderson impurity model is
formulated within the tight-binding approximation for determining
the host band-structure and the impurity-host hybridization. Using
the quantum Monte Carlo technique, we observe a finite local moment
for an N impurity, and  long-range ferromagnetic correlations
between two N impurities. The ferromagnetic correlations are
strongly influenced by the impurity bound state. When the
ferromagnetic correlation between a pair of impurities is mapped
onto the isotropic Heisenberg model for two spin-$1/2$ particles,
the effective exchange constant $J_{12}$ is found to increase with
increasing temperature. Similar temperature dependence of $J_{12}$
is also obtained in other diluted magnetic semiconductors, such as
zincblende ZnO doped with Mn. The temperature dependence of $J_{12}$
suggests that the mapping of the full Hamiltonian onto the spin
Hamiltonian cannot fully describe the magnetic correlations for the
diluted magnetic semiconductors at least in the limit of low
impurity spin.
\end{abstract}

\pacs{75.50.Pp, 75.30.Hx, 75.40.Mg} \maketitle

\section{Introduction}
There are  currently a  large number of observations of
ferromagnetism associated with doping of {\it a priori}
non-magnetic species into non-magnetic oxide semi-conductors
\cite{CoeyNature, Coey2, CoeyHfO2}. The subject has  been given
the name of d$^0$ magnetism to emphasize the fact that the
magnetism is probably not coming from partially filled d-orbitals,
but from  moments induced in the p-orbitals of the oxygen band
\cite{Stoneham,Sawatzky,PemmarajuSanvito}. Even in cases where
there are  partially filled d-orbitals, discrepancies between bulk
measurements indicating ferromagnetism, and microscopic
measurements by XMCD finding paramagnetic transition metal ions,
have suggested that the ferromagnetism comes from vacancies in the
oxygen lattices \cite{Tietze}. While a fertile area of current
investigation, the subject remains obscure because of problems of
irreproducibility for different samples, sample inhomogeneity,
stability in time, {\it etc.}, and the exact origin of the
magnetism is not yet established. This is unlike the case of
regular diluted magnetic semi-conductors such as (Ga,Mn)As
\cite{Ohno}, where ferromagnetism is reproducible in controlled
samples. The theoretical situation is also much less clear for
d$^0$ magnetism than in the case of ``classic'' diluted magnetic
semiconductors where the origin of the magnetic moments, the
substituted magnetic ion, is in no doubt and, in the case of
Mn$^{2+}$, quite large (S=5/2). Magnetic exchange interactions
leading to ferromagnetism are mediated by holes in the p band. For
(Ga,Mn)As, Curie temperatures can be estimated with reasonable
accuracy, starting from band structure to first derive the
effective exchange at different distances between spins in  an
effective spin Hamiltonian \cite{Liechtenstein} which is then
studied via classical Monte Carlo or other approximate methods
\cite{Bergqvist,Sato2004,Bouzerar2}. Although this approach  is
adequate for bulk properties such as the ordering temperature for
a variety of materials, fuller treatment of the effects of the
correlation are needed for understanding more detailed properties
such as the local densities of states and impurity band structure
\cite{Oheetal}. Even for thermodynamic properties such as the
Curie temperature, there is good reason to doubt straightforward
application of methods apparently working for ``classic'' diluted
magnetic semiconductors to the materials that may show  d$^0$
magnetism. In this case, the induced moments postulated  in the
oxygen bands are not fully  localized, but are formed by the same
holes that  generate  effective exchange. It was recently argued,
for example,  that \cite{Droghetti} because of this, exchanges
estimated from local approximations, such as the  local spin
density approximation (LSDA) may consistently over-estimate the
tendency for ferromagnetism and even formation of local moments,
compared to extensions, such as self-interaction correction (SIC).
Thus much of the theoretical literature may be over-optimistic in
predictions of ferromagnetism, even at zero temperature. One can
even question whether the approach is valid to  first project  the
full Hamiltonian onto the zero-temperature spin Hamiltonian, to
which the effects of thermal fluctuations are added.
\par
It is  important, therefore,  to develop methods which treat
correlations correctly and that do not rely on arbitrary
approximations or on the separation between spin and charge
fluctuations. Such a method has been introduced and applied to
models of the ``classic'' diluted magnetic semiconductors
\cite{Bulut,Tomoda,ZnO}, namely  Quantum Monte Carlo
methods based on the Hirsch-Fye algorithm.
These can directly give spin correlations at finite
temperatures without making any assumption of projection onto an
effective Hamiltonian. In this paper we apply the method to an
interesting case of possible d$^0$ magnetism, namely MgO diluted
with nitrogen. Our aim is then to answer, by means of Quantum Monte
Carlo methods the following questions. Firstly,   does such an
unbiased calculation predict ferromagnetic correlations in such a
material? Secondly, of more general interest, do the  standard
approaches, as  described above, still apply? In fact ferromagnetism
has previously been predicted at low concentrations \cite{Katayama}
in the doped alkaline-earth metal oxides (MgO, CaO, BaO, SrO) doped
with N and C. Our results are not completely comparable to those
results, in that we shall consider a simpler model Hamiltonian
 for the host and  two impurities only, rather than a finite concentration.
We should be able to draw useful general conclusions both for the particular
material and the methods.

\section{Ferromagnetic correlations between two N impurities}
\subsection{Impurity model}

In order to describe N impurities in MgO host, we take the two-step
calculations. Firstly, the Haldane-Anderson impurity model
\cite{Haldane} is formulated within the tight-binding approximation
for determining the host band structure and the impurity-host
hybridization. Secondly, the magnetic correlations of the
Haldane-Anderson impurity model at finite temperatures are
calculated by the Hirsch-Fye quantum Monte Carlo technique
\cite{QMC}.

The Haldane-Anderson impurity model is defined as
\begin{eqnarray}
  H &=&
  \sum_{\textbf{k},\alpha,\sigma}[\epsilon_{\alpha}(\textbf{k})-\mu]
  c^{\dag}_{\textbf{k}\alpha\sigma}c_{\textbf{k}\alpha\sigma} \notag\\
  &+&\sum_{\textbf{k},\alpha,\textbf{i},\xi,\sigma}(V_{\textbf{i}\xi\textbf{k}\alpha }
 p^{\dag}_{\textbf{i}\xi\sigma} c_{\textbf{k}\alpha\sigma}
   + H.c.) \notag\\ &+ &(\epsilon_p-\mu)\sum_{\textbf{i},\xi,\sigma}
   p^{\dag}_{\textbf{i}\xi\sigma}p_{\textbf{i}\xi\sigma}
   + U\sum_{\textbf{i},\xi}n^{\dag}_{\textbf{i}\xi\uparrow}n_{\textbf{i}\xi\downarrow},
   \label{E-Ham}
\end{eqnarray}
where $c^{\dag}_{\textbf{k}\alpha\sigma}$
($c_{\textbf{k}\alpha\sigma}$) is the creation (annihilation)
operator for a host electron with wavevector $\textbf{k}$ and spin
$\sigma$ in the valence ($\alpha = v$) or conduction ($\alpha = c$)
band, and $p^{\dag}_{\textbf{i}\xi\sigma}$
($p_{\textbf{i}\xi\sigma}$) is the creation (annihilation) operator
for a localized electron at impurity site $\textbf{i}$ in orbital
$\xi$ ($\xi$ = $x$, $y$, $z$) and spin $\sigma$ with
$n_{\textbf{i}\xi\sigma}=p^{\dag}_{\textbf{i}\xi\sigma}p_{\textbf{i}\xi\sigma}$.
Here, $\epsilon_{\alpha}(\textbf{k})$ is the host band dispersion,
$\mu$ the chemical potential, $V_{\textbf{i}\xi\textbf{k}\alpha}$
the mixing between the impurity and host, $\epsilon_p$ the impurity
$2p$ orbital energy, and $U$ the on-site Coulomb repulsion for the
impurity.

The energy bands $\epsilon_{\alpha}(\textbf{k})$ for MgO host, and
the impurity-host hybridization $V_{\textbf{i}\xi\textbf{k}\alpha}$
will be calculated within the tight-binding approximation. For a
large number of simple oxides, the on-site Coulomb repulsion energy
of holes in an oxygen p orbital is 5-7 eV \cite{Yunoki}. For the
on-site Coulomb repulsion of $2p$-orbitals at an N impurity site in
the compound MgO, the experimental value is unknown, so here we take
it as $U$ = 6 eV. In addition, the experimental value of impurity
$2p$ energy $\epsilon_p$(N) in Mg(O,N) is also unknown. In the
following, we use $\epsilon_p$(N) as the parameter satisfying
$\epsilon_p$(N) $>$ $\epsilon_p$(O)= -2.03 eV, where the value of
$\epsilon_p$(O) is taken from the tight-binding parameters for the
compound MgO (see below).

For the compound Mg(O,N), there is one $2p$ hole at an N$^{2-}$
site, so it is reasonable to neglect the Hund couplings among
different $2p$ orbitals at an N$^{2-}$ site.

\subsection{Tight-binding approach for the MgO band structure and
the N-MgO hybridization}

\begin{figure}[tbp]
\includegraphics[width = 8.5 cm]{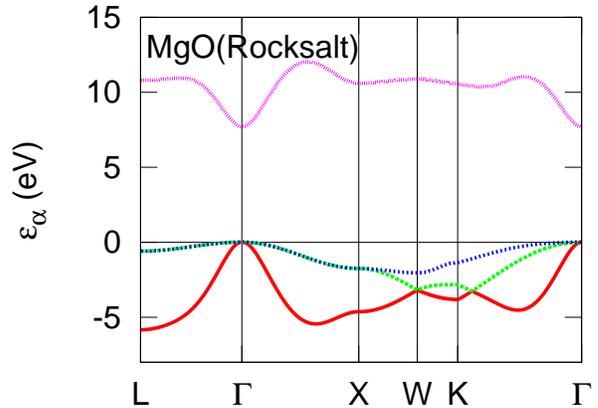}
\caption{(Color online) Energy bands of MgO with rocksalt
crystal structure. These results were reproduced using
the tight-binding parameters in Ref.~\cite{MgO}, where one $3s$ orbital
of Mg and three $2p$ orbitals of O are included.} \label{F-Eig}
\end{figure}

In this section, we discuss the tight-binding calculation for the
MgO band-structure, and the hybridization between an N impurity
and MgO host. For MgO with rocksalt structure, the band structure
$\epsilon_{\alpha}(\textbf{k})$ had already been calculated with a
set of tight-binding parameters \cite{MgO}. In this approach, the
basis consists of a $3s$ orbital for the cation Mg$^{2+}$ and
three degenerate $2p$ orbitals for the anion O$^{2-}$. The orbital
energies are $\epsilon_s$(Mg) = 9.88 eV, $\epsilon_p$(O)= -2.03
eV. In addition, the mixing values between the $s$ orbital of
Mg$^{2+}$ and the $p$ orbitals of O$^{2-}$ up to the 3rd-nearest
neighbors are taken to be $(sp\sigma)_1$ = 1.10 eV, $(ss\sigma)_2$
= -0.18 eV, $(pp\sigma)_2$ = 0.65 eV, $(pp\pi)_2$ = -0.07 eV, and
$(sp\sigma)_3$ = 0.89 eV, where $()_n$ means the mixing of the
$n$-th nearest neighbors. Using these tight-binding parameters, we
have reproduced the band structure of rocksalt MgO as shown in
Fig. \ref{F-Eig}, where band structure consists of one conduction
band and three valence bands. The conduction band mainly comes
from $3s$ orbital of Mg, and valence bands mainly come from $2p$
orbitals of O. The top of valence bands and the bottom of
conduction band are located at the $\Gamma$ point with a direct
gap of 7.72 eV.

Next, we discuss the calculation of the hybridization between N impurity
and MgO host within the tight-binding approximation.
The hybridization matrix element
$V_{\textbf{i}\xi\textbf{k}\alpha}\equiv\langle\varphi_{\xi}
(\textbf{i})|H|\Psi_{\alpha}(\textbf{k})\rangle$ has the form of
\begin{eqnarray}
V_{\textbf{i}\xi\textbf{k}\alpha } &=& \frac{1}{\sqrt{N}}e^{i
\textbf{k}\cdot \textbf{i}}\sum_{o,\textbf{n}}e^{i \textbf{k}\cdot
(\textbf{n}-\textbf{i})}a_{\alpha o}(\textbf{k})
\langle\varphi_{\xi}(\textbf{i})|H|\varphi_{o}(\textbf{n})\rangle
\notag\\
& \equiv &\frac{1}{\sqrt{N}}e^{i \textbf{k}\cdot
\textbf{i}}V_{\xi\alpha }(\textbf{k}),\label{E-Mix}
\end{eqnarray}
where $\varphi_{\xi} (\textbf{i})$ is the impurity $2p$-state
($\xi =x, y,z$) at
site $\textbf{i}$, and $\Psi_{\alpha}(\textbf{k})$ is the host state
with wavevector $\textbf{k}$ and band index $\alpha$, which is
expanded by atomic orbitals $\varphi_{o}(\textbf{n})$ with orbital
index $o$ and site index $\textbf{n}$. Here, $N$ is the total number
of host lattice sites, and $a_{\alpha o}(\textbf{k})$ is an
expansion coefficient. For the mixing integrals of
$\langle\varphi_{\xi}(\textbf{i})|H|\varphi_{o}(\textbf{n})\rangle
$, $\xi$ denotes the three $2p$ orbitals of N$^{2-}$, and $o$
represents the $3s$ orbital of Mg$^{2+}$ and $2p$ orbitals
of O$^{2-}$. As shown by Slater and Koster \cite{TBSK}, these mixing
integrals up to the 3rd-nearest neighbors can be expressed by four
integrals $(sp\sigma)_1$ $(pp\sigma)_2$, $(pp\pi)_2$, $(sp\sigma)_3$
and direction cosines $l$, $m$ and $n$ in the two-center
approximation, where $()_n$ means the mixing of the $n$-th nearest
neighbors. As the experimental values of the above four integrals
are unknown, here we take these mixing integrals between N and MgO
as the same mixing values between O and MgO in Ref.\cite{MgO}.
Thus, we have $(sp\sigma)_1$ = 1.10 eV,
$(pp\sigma)_2$ = 0.65 eV, $(pp\pi)_2$ = -0.07 eV, and $(sp\sigma)_3$
= 0.89 eV for the impurity-host hybridization.

\begin{figure}[tbp]
\includegraphics[width = 8.5 cm]{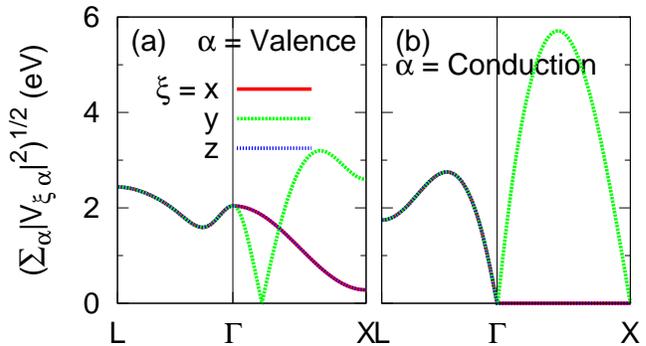}
\caption{ (Color online) Hybridization function
$\overline{V}_{\xi}(\textbf{k})$ between $\xi$
orbital of an N impurity and MgO host (a) valence bands or (b) conduction band.
Here, the mixing parameters between N and MgO are taken as
the same mixing values between O and MgO in Ref. \cite{MgO}.}
\label{F-Mix}
\end{figure}

Figure \ref{F-Mix} displays results on the impurity-host mixing
function $\overline{V}_{\xi}(\textbf{k})$ defined by
\begin{equation}
\overline{V}_{\xi}(\textbf{k})  \equiv \big(
\sum_{\alpha}|V_{\xi\alpha}(\textbf{k})|^2 \big)^{1/2}
\label{E-Vbar}
\end{equation}
where $\xi$ is a $2p(x,y,z)$ orbital of an N impurity. In
Eq.~(\ref{E-Vbar}), the summation over $\alpha$ is performed only
over the valence bands (Fig. \ref{F-Mix}(a)) or the conduction
band (Fig. \ref{F-Mix}(b)). Here, $\overline{V}_{\xi}(\textbf{k})$
is plotted along various directions in the Brillouin zone for
rocksalt crystal structure. It is observed that, at the $\Gamma$
point, the total hybridization between the $\xi$ orbital of an N
impurity and the MgO valence bands is finite, while that with the
MgO conduction band is zero. For the host MgO, the gap edge is
located at the $\Gamma$ point, hence the value of $\overline{V}$
near $\Gamma$ will be particularly important in determining the
energy of the impurity bound state, which may appear in the gap
due to the mixing between impurity and host.

\subsection{Quantum Monte Carlo results}

In this section, we present results on the magnetic correlations of
the Haldane-Anderson impurity model obtained by the Hirsch-Fye
quantum Monte Carlo (QMC) technique \cite{QMC}. The parameters
related to the MgO host and N impurity have been calculated within
the tight-binding approach described above. The following results
were obtained with more than 10$^{5}$ Monte Carlo sweeps and
Matsubara time step $\Delta\tau=0.225$.

\begin{figure}[tbp]
\includegraphics[width = 8.5 cm]{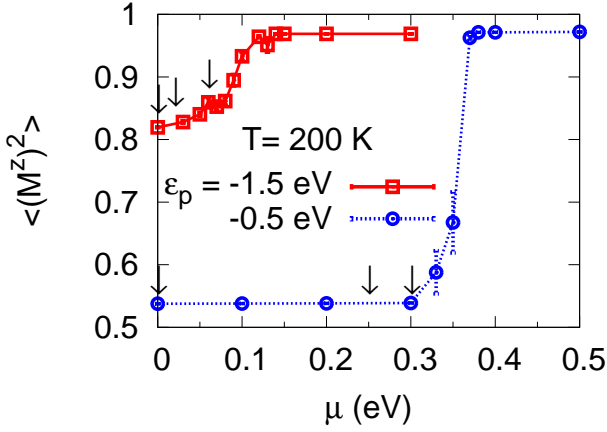}
\caption{ (Color online) Square of magnetic moment
$\langle(M^z)^2\rangle$ of an N impurity versus chemical potential
$\mu$ at $T=200K$. The experimental value of impurity $2p$ energy
$\epsilon_p(N)$ in Mg(O,N) is unknown, but it should satisfy the
relation $\epsilon_p(N)$ $> \epsilon_p(O)$ = $-2.03$ eV. Here, the
arrows indicate the different position of $\mu$ investigated by Fig.
\ref{F-C2}.} \label{F-C1}
\end{figure}

We first discuss the local moment formation for an N impurity
in the MgO host. For this purpose, we have performed QMC simulations to
calculate $\langle (M^z)^2 \rangle$, where
\begin{equation}
M^z = n_{\textbf{i}\xi\uparrow} - n_{\textbf{i}\xi\downarrow}
\end{equation}
is the magnetization operator for a $\xi$ orbital of an N impurity
at site $\textbf{i}$. Fig. \ref{F-C1} shows $\langle (M^z)^2
\rangle$ versus the chemical potential $\mu$ at temperature $T = 200
K$, where $0 < \mu < 0.5$eV. As mentioned before, the experimental
value of impurity $2p$ energy $\epsilon_p(N)$ in Mg(O,N) is not
known, so it is taken as a parameter satisfying the relation
$\epsilon_p(N)$ $> \epsilon_p(O)$ = $-2.03$ eV. As shown in Fig.
\ref{F-C1}, the sharp increases in the magnitude of the magnetic
moment are observed around $0.08$eV and $0.35$eV for the impurity
$2p$ energy $\epsilon_p=-1.5$eV and $\epsilon_p=-0.5$eV,
respectively. In our model $2p$ orbitals are degenerate, and thus
the calculated curves in Fig. \ref{F-C1} do not change with $\xi=x$,
$y$ or $z$. In addition, no such sharp increases (or sharp
decreases) are observed near the bottom of the conduction band. As
shown in Fig. \ref{F-Mix}, the values of the hybridization with
bottom of conduction band are around zero and thus much weaker than
those with top of valence bands.

According to the Hartee-Fock \cite{2DFM1} and QMC calculations
\cite{Bulut,Tomoda,ZnO}, the presence of a sharp increase (or a
sharp decrease) in $\langle (M^z)^2\rangle$ versus $\mu$ implies the
existence of an impurity bound state (IBS) at this energy, and the
IBS plays an important role in determining the strength of the
ferromagnetic (FM) correlations. When the IBS is unoccupied,  FM
correlations can develop between the impurities. When the IBS is
occupied, the FM correlations become weaker.

When two N impurities are introduced in the MgO host, they can
replace any two O positions in the rocksalt lattice
MgO.  Let us then briefly explain how our QMC calculations of the
magnetic correlations function $\langle M^z_1M^z_2\rangle$
actually proceed with the different spatial positions of the two
impurities. When the position of the two impurities changes, the
corresponding hybridization $V_{\textbf{i}\xi \textbf{k}\alpha}$,
including the impurity spatial position $\textbf{i}$ as defined in
Eq.(2), will change; the Green's function between impurity 1 and
impurity 2, including the factor $V^{\ast}_{1\xi
\textbf{k}\alpha}V_{2\xi \textbf{k}\alpha}$, will change; and thus
the magnetic correlations function $\langle M^z_1M^z_2\rangle$
changes. See Ref.\cite{QMC2} for more details of the QMC
calculations.

\begin{figure}[tbp]
\includegraphics[width = 8.5 cm]{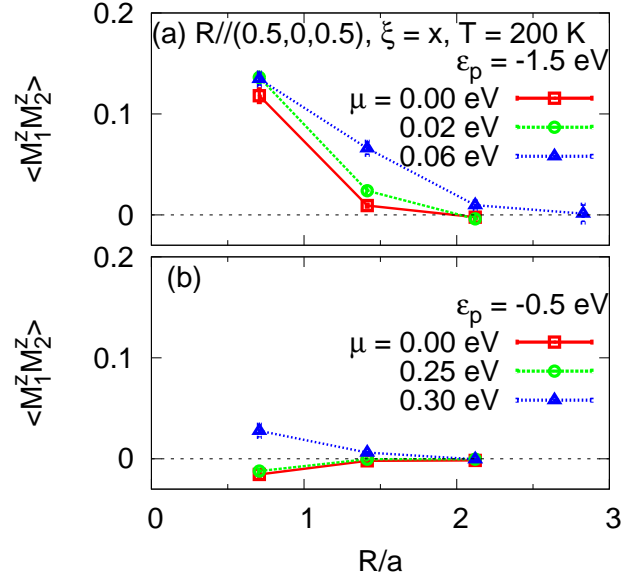}
\caption{ (Color online) Magnetic correlation function $\langle
M^z_{1}M^z_{2}\rangle$ between $\xi=x$ orbitals of the N impurities
versus distance $R/a$ for (a) the impurity $2p$ energy $\epsilon_p$
= -$1.5$ eV and (b) $\epsilon_p$ = -$0.5$ eV at $T = 200 K$.
Direction $R//(0.5,0,0.5)$ is along one of the 12 nearest N-N
neighbors in rocksalt structure. } \label{F-C2}
\end{figure}

The magnetic correlation function $\langle M^z_{1}M^z_{2}\rangle$
between $\xi=x$ orbitals of the impurities versus the impurity
separation $R/a$ at temperature $T = 200 K$ is shown in Fig.
\ref{F-C2}. The direction $R//(0.5,0,0.5)$ is chosen to be along one
of the 12 nearest N-N neighbors in rocksalt structure, and $a$ is
the lattice constant. Results in Fig. \ref{F-C2}(a) are obtained
with fixed impurity energy $\epsilon_p=-1.5$ eV for various chemical
potential $\mu$. The impurity spins exhibit FM correlations at
chemical potential $\mu=0.0$ eV. It is noted that the IBS of Mg(O,N)
lies near 0.08 eV in the gap, close to the top of the valence band
as displayed in Fig. \ref{F-C1}. By increasing $\mu$ to 0.06 eV, the
FM correlations become larger, and the range of the FM correlations
becomes longer. Further increasing $\mu$, the FM correlations
become weaker. This is because the IBS of Mg(O,N) becomes occupied
when $\mu$ is increased to above 0.06 eV, as seen in Fig.
\ref{F-C1}. Results in Fig. \ref{F-C2}(b) are obtained with fixed
impurity energy $\epsilon_p=-0.5$ eV for various values of the
chemical potential $\mu$. The impurity spins exhibit quite weak
antiferromagnetic (AFM) correlations due to the superexchange
interaction at $\mu=0.0$ eV. It is worth pointing out that the IBS
of Mg(O,N) lies around 0.35 eV, deep in the band gap as shown in
Fig. \ref{F-C1}. Increasing $\mu$ to 0.3 eV, the FM correlations
appear with stronger magnitude and longer range. Further increasing
$\mu$ to above 0.3 eV, the FM correlations become weaker. This is
because the IBS of Mg(O,N) is occupied when $\mu
> 0.3$ eV, as displayed in Fig. \ref{F-C1}.

\begin{figure}[tbp]
\includegraphics[width = 8.5 cm]{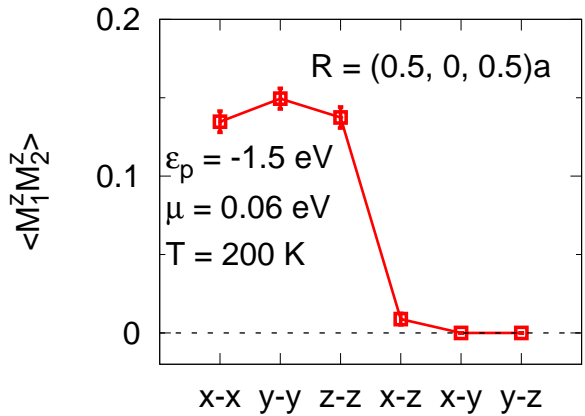}
\caption{ (Color online) Magnetic correlation function
$\langle M^z_{1}M^z_{2}\rangle$ between $2p$ orbitals
of two N impurities. Distance $R = (0.5, 0, 0.5)a$
is one of the 12 nearest N-N neighbors in rocksalt structure.
Here,  $\epsilon_p$ = -$1.5$ eV, $\mu$ = $0.06$ eV
and $T = 200 K$.} \label{F-Anyorb}
\end{figure}

In the direction $R//(0.5,0,0.5)$, let us discuss the correlation
function $\langle M^z_{1}M^z_{2}\rangle$ between other orbitals of
two N impurities in the MgO host. Considering the symmetry in the
direction $R//(0.5,0,0.5)$, $\langle M^z_{1}M^z_{2}\rangle$ between
$\xi=z$ orbitals of two N impurities have the same value between
$\xi=x$ orbitals of two N impurities. For the $\langle
M^z_{1}M^z_{2}\rangle$ between $\xi=y$ orbitals of two N impurities
in the direction $R//(0.5,0,0.5)$, our QMC calculations (not present
here) show that it is short-range FM correlation. For the nearest
N-N neighbors distance $R = (0.5, 0, 0.5)a$ in the case that
$\epsilon_p$ = -$1.5$ eV, $\mu$ = $0.06$ eV and $T = 200 K$, the
magnetic correlation function $\langle M^z_{1}M^z_{2}\rangle$
between different $\xi$ orbitals is shown in Fig. \ref{F-Anyorb}. It
is found that the magnetic correlations between different orbitals
of two N impurities are much smaller than the values between the same
orbitals.

For the two N impurities in the MgO host, there are 12 nearest N-N
neighbors in rocksalt structure : $(0, \pm 0.5,\pm 0.5)a$, $(\pm
0.5,0,\pm 0.5)a$, $(\pm 0.5,\pm 0.5, 0)a$. For the direction $R$
along other 11 nearest N-N neighbors, $\langle
M^z_{1}M^z_{2}\rangle$ of two N impurities could be got by symmetry
without calculation. For example, $\langle M^z_{1}M^z_{2}\rangle$
between $\xi=x$ orbital of two N impurities in direction
$R//(0,0.5,0.5)$ is the same value as that between $\xi=y$ orbitals
of two N impurities in direction $R//(0.5,0,0.5)$. Thus, for
$\epsilon_p$ = -$1.5$ eV and $\mu$ = $0.06$ eV, the long-range FM
correlation between two N impurities could be observed at $T = 200
K$ along all of the 12 nearest N-N neighbors in rocksalt Mg(O,N).

The calculations presented above do not include the Jahn-Teller(JT)
distortion, which might occur in the compound Mg(O,N). As the JT
distortion is included, the energy level of x, y, z orbitals of the
impurity N will differ, being higher for some orbitals and being
lower for the others, but the values changed are often quite small.
As a result, the IBS shown in Fig. \ref{F-C1} will differ for
different p orbitals, being shallower in the gap for those having
lower energy level and being deeper in the gap for those having
higher energy level, and again the shift values of the IBS will be
quite small. More importantly, the long-range FM correlations, as
shown in Fig.\ref{F-C2}, will become stronger for those having the
shallower IBS and become weaker for those having the deeper IBS, and
of course the change will be quite small. So, here we argue that the
JT distortion will play a quite small role in our calculations.

\section{Mapping of Ferromagnetism onto a Heisenberg model}

The ferromagnetic correlation between two N impurities is mapped
onto the isotropic Heisenberg model for two spin-$1/2$ particles
\begin{equation}\label{E-Heisenberg}
    H = -J_{12}\textbf{M}_1\cdot \textbf{M}_2.
\end{equation}
At finite temperature, defining $\beta=1/k_B T$ with $k_B$ the
Boltzmann constant, the impurity-impurity correlation is
defined as
\begin{equation*}
\langle M_1^zM_2^z\rangle = Tr(M_1^zM_2^ze^{-\beta H})/Tr(e^{-\beta
H}).
\end{equation*}
Considering
\begin{eqnarray*}
 \textbf{M}_1\cdot \textbf{M}_2&=&
 \frac{1}{2}[(\textbf{M}_1+\textbf{M}_2)^2-\textbf{M}^2_1-\textbf{M}^2_2]\\
   &=&\frac{1}{2}[S(S+1)-2s(s+1)],
\end{eqnarray*}
where $s=1/2$, $S=0, 1$, and the trace here could be taken as
\begin{eqnarray*}
    Tr(\cdots)=\sum^{S}_{S^z=-S}\sum^{1}_{S=0}(\cdots)
    =\sum^{1/2}_{s_2^z=-1/2}\sum^{1/2}_{s_1^z=-1/2}(\cdots),
\end{eqnarray*}
we have
\begin{equation*}
    \langle M_1^zM_2^z\ \rangle =
    \frac{1}{4}\cdot\frac{1-e^{-\beta J_{12}}}{3+e^{-\beta J_{12}}},
\end{equation*}
where the unit of $\langle M_1^zM_2^z\rangle$ is
$1^2=(2s)^2=(2\mu_B)^2$, $\mu_B$ is Bohr magneton. To be consistent
with our QMC calculation results, whose unit of $\langle
M_1^zM_2^z\rangle$ is $\mu_B^2$, the above equation is modified as
\begin{equation}\label{E-Coupling}
    \langle M_1^zM_2^z\ \rangle = \frac{1-e^{-\beta J_{12}}}{3+e^{-\beta
    J_{12}}}.
\end{equation}
Thus, inverting this relation, we can deduce an effective exchange coupling $J_{12}$ between two N
impurities from
\begin{equation}\label{E-J12}
    J_{12} = k_BT\ln\frac{1+\langle M_1^zM_2^z\rangle}{1-3\langle
    M_1^zM_2^z\rangle}.
\end{equation}

\begin{figure}[tbp]
\includegraphics[width = 8.5 cm]{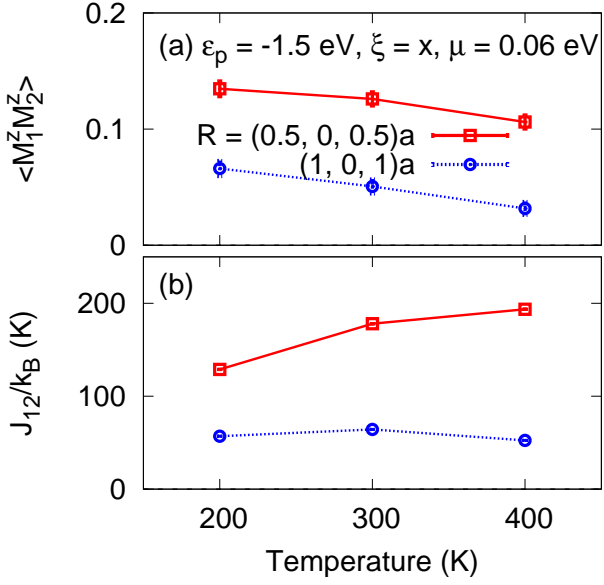}
\caption{ (Color online) Temperature dependence of (a)the  magnetic
correlation function $\langle M^z_{1}M^z_{2}\rangle$ between $x$
orbitals of two N impurities, and (b) corresponding exchange
coupling $J_{12}$ calculated by Eq.(\ref{E-J12}). Here, $\epsilon_p$
= -$1.5$ eV, $\mu$ = 0.06 eV, $R=(0.5,0,0.5)a$ or $R=(1,0,1)a$.}
\label{F-JMgO}
\end{figure}

Figure \ref{F-JMgO} shows the temperature dependence of the
magnetic correlation function $\langle M^z_{1}M^z_{2}\rangle$ and
the corresponding exchange coupling $J_{12}$ between $\xi=x$
orbitals of two N impurities. Here, the following parameters are
taken as $\epsilon_p$ = -$1.5$ eV, $\mu$ = $0.06$ eV,
$R=(0.5,0,0.5)a$ or $R=(1,0,1)a$. For the two N impurities with
distance $R=(0.5,0,0.5)a$, it is found that with increasing
temperature from $200 K$ to $400 K$, magnetic correlation function
$\langle M^z_{1}M^z_{2}\rangle$ decreases, but the exchange
coupling $J_{12}$ increases. For the two N impurities with
distance $R=(1,0,1)a$, the temperature dependence of exchange
coupling $J_{12}$ becomes weaker.

For $\langle M^z_{1}M^z_{2}\rangle$ and $J_{12}$ between $\xi=y$ or
$z$ orbitals of two N impurities in the direction $R//(0.5,0,0.5)$,
the similar behaviors are observed. For the direction $R$ along
other 11 nearest N-N neighbors, the behaviors of temperature
dependence could be found by symmetry without calculation as
discussed in the last section. Thus, with $\epsilon_p$ = -$1.5$ eV
and $\mu$ = $0.06$ eV, the temperature dependence of $\langle
M^z_{1}M^z_{2}\rangle$ and $J_{12}$ between two N impurities, as
shown in Fig. \ref{F-JMgO}, could be observed along all of the 12
nearest N-N neighbors in rocksalt Mg(O,N).

\section{Discussion}

\begin{figure}[tbp]
\includegraphics[width = 8.5 cm]{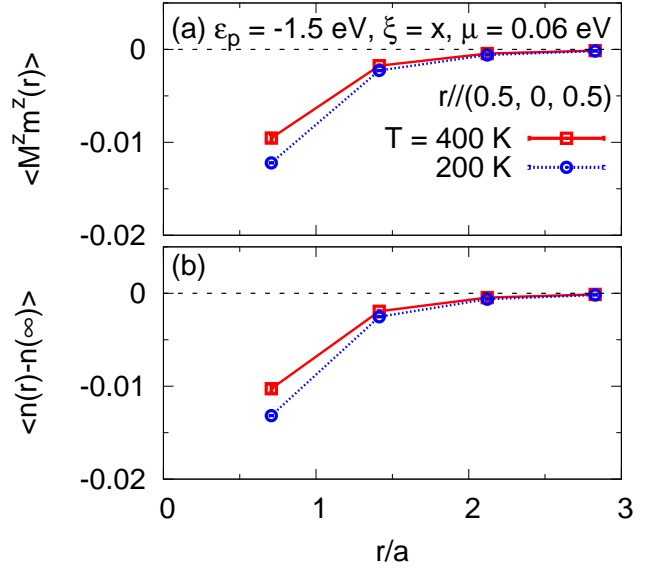}
\caption{ (Color online) Spatial variation  of
(a) the magnetic correlation function $\langle M^zm^z(r)\rangle$
between $x$ orbital of an N impurity and MgO host, and
(b) the number of polarized host electrons $\langle n(r)-n(\infty)\rangle$.
Here, $\epsilon_p$ = -$1.5$ eV, $\mu$ = $0.06$ eV, and
$r//(0.5,0,0.5)$. } \label{F-DCK}
\end{figure}

To understand the long-range FM correlation function $\langle
M^z_{1}M^z_{2}\rangle$ between two N impurities in MgO host, as
shown in Fig. \ref{F-C2}, which is mediated by the impurity-induced
polarization of the host electron spins, it is more convenient to
study the local density of states in real space around the impurity.
Here we have considered the impurity-host correlation function
$\langle M^zm^z(\textbf {r})\rangle$ and the number of polarized
host electrons $\langle n(\textbf {r})-n(\infty)\rangle$ for the
case of one-impurity N in the MgO host. Here, $\textbf {r}$ is the
site of host electron and the impurity N is located at site $\textbf
{r}$ = 0. In addition, $n(\infty)$ means the number of host
electrons at infinity distance, thus $\langle n(\textbf
{r})-n(\infty)\rangle$ is negative and represents the number of
holes. The magnetization $m^z(\textbf {r})$ and number
  $n(\textbf {r})$ operators are defined as
\begin{eqnarray}
m^z(\textbf {r})&=&\sum_{\alpha}( n_{\alpha \textbf {r}\uparrow}
-n_{\alpha \textbf {r}\downarrow}),\\
n(\textbf {r})&=&\sum_{\alpha}( n_{\alpha \textbf {r}\uparrow}
+n_{\alpha \textbf {r}\downarrow}),
\end{eqnarray}
where $n_{\alpha\textbf{r}\sigma}=c^{\dag}_{\alpha\textbf{r}\sigma}
c_{\alpha\textbf{r}\sigma}$ is the number operator for host
electrons with band index $\alpha$ and site $\textbf {r}$ and spin
$\sigma$. With impurity $2p$ energy $\epsilon_p$ = -$1.5$ eV and
chemical potential $\mu$ = 0.06 eV and direction $\textbf
{r}//(0.5,0,0.5)$, the long-range antiferromagnetic (AFM)
correlation between $x$ orbital of N impurity and MgO host is
observed in Fig. \ref{F-DCK}(a), and the polarized host electrons
with long-range distribution are also observed in Fig.
\ref{F-DCK}(b). As the holes in the valence band round the impurity
N are spin-polarized, the valence band is spin-polarized. Comparing
Fig. \ref{F-C2}(a) with Fig. \ref{F-DCK}(a), it is confirmed that
the long-range AFM impurity-host correlations contribute to the
long-range FM impurity-impurity correlations. With increasing
temperature, as shown in Fig. \ref{F-DCK}(a), the magnitude of AFM
impurity-host correlation function $\langle M^zm^z(\textbf
{r})\rangle$ decreases, which induces the decreasing FM
impurity-impurity correlation function $\langle
M^z_{1}M^z_{2}\rangle$ with increasing temperature as shown in Fig.
\ref{F-JMgO}(a).

\begin{figure}[tbp]
\includegraphics[width = 8.5 cm]{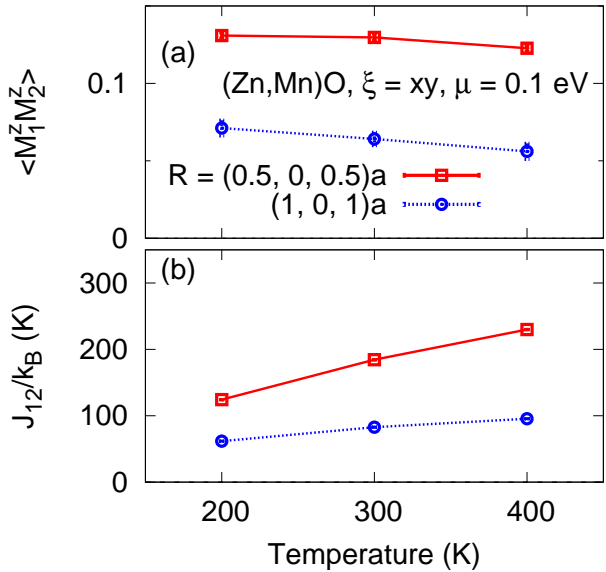}
\caption{ (Color online) For zincblende ZnO doped with Mn,
temperature dependence of (a) the magnetic correlation function $\langle
M^z_{1}M^z_{2}\rangle$ between $xy$ orbitals of two Mn impurities,
and (b) the corresponding effective exchange coupling $J_{12}$ calculated by Eq.
(\ref{E-J12}). Here $\epsilon_d$ = -U/2 + $\mu$, $\mu$ = 0.1 eV, and
$R//(0.5,0,0.5)$. } \label{F-JZnO}
\end{figure}

To understand the temperature dependent exchange coupling $J_{12}$
shown in Fig. \ref{F-JMgO}(b), we have studied a ``classic" diluted
magnetic semiconductor, zincblende ZnO doped with Mn. As shown in
Fig. \ref{F-JZnO}, a similar temperature-dependent exchange coupling
$J_{12}$ is obtained. Here, magnetic correlation function
$\langle M^z_{1}M^z_{2}\rangle$
between $xy$ orbitals of two Mn impurities is
calculated, and the corresponding exchange coupling $J_{12}$ is
given by Eq.(\ref{E-J12}). Because the experimental value of
$\epsilon_d$ for Mn in ZnO host is unknown, here we use the
symmetric case of $\epsilon_d$ = -U/2 + $\mu$ so that the impurity
sites develop large magnetic moments. For the compound (Zn,Mn)O, the
value of the on-site Coulomb repulsion for Mn$^{2+}$ is taken as
$U=5.2$eV by comparing with the photoemission spectroscopy
measurements \cite{MnU}. The chemical potential value $\mu$ = 0.1 eV
is set close to the impurity bound state. The direction
$R//(0.5,0,0.5)$ is chosen to be along one of the 12 nearest Mn-Mn
neighbors in zincblende structure. We find that the exchange
constant $J_{12}$ increases with increasing temperature as was the
case in Mg(O,N).

We have to note that our result of $J_{12}$ does indeed
contradict the common assumption that people make when using the
Lichtenstein formula, but here we are dealing with a low spin situation,
in contrast to what is usually (but not always) done in the literature.
Thus our results are most pertinent for d$^0$ situations, and
the results for zincblende ZnO doped with Mn is in fact based
on an effective spin-$1/2$ model.

\section{Summary and Conclusions}

In summary, we have studied  possible d$^0$ ferromagnetism for the
compound Mg(O,N) in the dilute impurity limit based on the
Haldane-Anderson impurity model. The band structure of the MgO host
were calculated using the tight-binding approach. The mixing
parameters between N and MgO are approximated as being the same as
between O and MgO. The QMC results show the development of a large
magnetic moment at an N impurity site, and  long-range ferromagnetic
correlations between two N impurities. The ferromagnetic correlation
between impurity pairs is mapped onto the isotropic Heisenberg model
for two spin-$1/2$ particles, and the effective exchange coupling
$J_{12}$ for given separation of impurities is found to increase
with increasing temperature. Similar temperature dependence of
$J_{12}$ is also obtained in ``classic" diluted magnetic
semiconductors, such as zincblende ZnO doped with Mn, suggesting
that the mapping is not fully valid even there at least in the limit
of low impurity spin. For the particular case of MgO doped with N
the results presented in this paper, in which no such approximation
is made and in which interactions are treated exactly, suggest that
there should be stable moments associated with each impurity and
that they should have ferromagnetic correlations. While the results
are for two impurities only, the long range of the correlations
suggests long-range ferromagnetic order for  low concentrations.

\section*{ACKNOWLEDGMENTS} This work was supported by the NAREGI
Nanoscience Project and a Grant-in Aid for Scientific Research
from the Ministry of Education, Culture, Sports, Science and
Technology of Japan, and NEDO. The authors thank the Supercomputer
Center at the Institute for Solid State Physics, University of
Tokyo, for the use of the facilities. T. Z. thanks the International
Frontier Center for Advanced Materials of Tohoku University and the
members of the IMR for their support of his stay in Sendai,
which enabled this collaboration. The authors acknowledge S. Parkin
for valuable discussion about the experiment of Mg(O,N).

\end{document}